\documentclass[conference]{IEEEtran}
\IEEEoverridecommandlockouts

\usepackage{cite}
\usepackage{amsmath,amssymb,amsfonts}
\usepackage{graphicx}
\usepackage{textcomp}
\usepackage{xcolor}
\usepackage{booktabs}
\usepackage{hyperref}

\usepackage{fancyhdr}
\hypersetup{hidelinks,pdftitle={Affective and Functional Needs in AI Use: User Patterns and Their Relation to Real-Life Support and Well-Being},pdfauthor={Jessy Wang-Sun, Estelle Yiyuan Sun, Hualin Zhang, Zihan Xu}}
\begin{document}
\thispagestyle{fancy}

\title{Affective and Functional Needs in AI Use: User Patterns and
Their Relation to Real-Life Support and Well-Being
\thanks{\copyright\ 2026 IEEE. Personal use of this material is permitted. Permission from IEEE must be obtained for all other uses, in any current or future media, including reprinting/republishing this material for advertising or promotional purposes, creating new collective works, for resale or redistribution to servers or lists, or reuse of any copyrighted component of this work in other works.}
}
\author{\IEEEauthorblockN{Jessy Wang-Sun}
\IEEEauthorblockA{\textit{Massachusetts Institute of Technology} \\
Cambridge, MA, USA \\
\textit{Boston University}, Boston, MA, USA \\
jwang33@mit.edu}
\and
\IEEEauthorblockN{Estelle Yiyuan Sun}
\IEEEauthorblockA{\textit{University of Florida} \\
Gainesville, FL, USA \\
yiyuansun@ufl.edu}
\and
\IEEEauthorblockN{Hualin Zhang}
\IEEEauthorblockA{\textit{Tsinghua University} \\
Beijing, China \\
Richard\_pku@163.com}
\and
\IEEEauthorblockN{Zihan Xu}
\IEEEauthorblockA{\textit{Northeastern University} \\
Boston, MA, USA \\
xu.zihan4@northeastern.edu}
}
\maketitle
\begin{center}
{\small Full paper accepted at the 14th International Conference on Affective Computing and Intelligent Interaction (ACII 2026), forthcoming in IEEE Xplore.}
\end{center}
\thispagestyle{fancy}
\begin{abstract}
Conversational AI systems are increasingly used to fulfill both functional and affective needs, yet how users perceive the AI-adoption effects varies. Existing research tends to treat these two needs as binary categories and to classify users by a single dominant need. However, this approach fails to capture users who engage with AI for both affective and functional purposes simultaneously, leaving it unclear how these needs combine in real-world AI use. To capture the multifaceted nature of users’ AI engagement, we developed the AI-Support measurement with 1,090 participants, grounded in social support theory, which evaluates four dimensions of perceived AI support: emotional/esteem, social companionship, informational, and instrumental support. This study provided (1) an AI-Support Scale developed to measure the perceived support types from human-AI interaction, (2) evidence that users might adopt AI with a high level of both functional and affective use, and (3) exploratory indicators of users’ real-life social support and well-being, enabling descriptive results aligning with theoretical expectations. Overall, this study highlights the importance of understanding individual differences in AI affordance use, enabling AI systems to be designed in alignment with different user adoption profiles rather than assuming uniform benefits across all users.
\end{abstract}

\section{Introduction}
In August 2025, OpenAI introduced GPT-5 as the default model for all ChatGPT users, removing access to earlier versions such as GPT-4o. This change triggered strong emotional reactions among users, with many reporting discomfort and dissatisfaction with the new system. Notably, users expressed a preference for the more "human-like" qualities of the previous model \cite{knight2025gpt5backlash}. OpenAI CEO Sam Altman acknowledged this response and emphasized the need for customizable AI personalities \cite{altman2025gpt5rollout}. This event highlights a broader issue: users do not engage with AI systems in uniform ways, and their affective reactions are shaped by underlying expectations and needs. 

Much existing scholarship on human–AI interaction has been grounded in the Computer as Social Actor (CASA) framework \cite{nass2000machines}, which suggests that users treat AI systems as social actors and apply human social norms during interaction. As a result, users may experience forms of social support from AI, such as emotional reassurance \cite{fulmer2018tess,chin2023support} or informational guidance \cite{skjuve2018chatbots}. These findings indicate that AI systems are not only used for functional purposes, but can also fulfill social and affective needs. Building on this perspective, prior research has begun to classify users based on their primary needs in AI interaction, with some users seeking emotional support and others preferring task-oriented assistance \cite{purington2017alexa,li2024love}. While these categorizations provide useful initial insights, they assume that users are primarily driven by a single dominant need. 

As the market increasingly shifts toward consolidating a wide range of user needs within a single AI platform, understanding how users perceive support from AI differently becomes a key factor in identifying users who may be more exposed to risk.  To understand human-AI interaction (HAII) from the perspective of users' needs, this study draws on social support theory as a structural lens. It conceptualizes the types of fundamental supports people need along multiple dimensions, including emotional, informational, instrumental, esteem, and companionship \cite{cutrona1992spouse,schwarzer2007roles}. Through this scope, we can analyze HAII with a close-up view of user heterogeneity, and understand how users navigate both affective and functional needs from AI. 
Hence, we examine \textbf{RQ1: How can users' affective and functional use of AI be measured?} and ask \textbf{RQ2: How are affective and functional needs jointly configured in AI use?}

The functional and affective needs defined in social support theory have long been recognized for their role in enhancing well-being within interpersonal relationships \cite{cohen1985buffering,chu2010socialsupport}, and emerging evidence suggests their relevance in human-AI interaction and relationships as well \cite{pan2025framework}. Users' engagement with AI may also be shaped by their real-life experiences.  Prior work suggests that individuals' social contexts can influence how they seek support through digital technologies \cite{hood2018online,zhang2025companions}. Hence, we extend the study result to an exploratory insight: \textbf{RQ3: \textbf{What are the levels of real-life social support and well-being among users with different profiles of affective and functional AI use?}}

To investigate the question, we conducted a survey study with 1,090 participants from China. We developed a survey measure of four AI-support dimensions (emotional, companionship, informational, and instrumental support) based on social support theory. We examined them as continuous dimensions of conversational AI use. Exploratory cluster analysis was also used to illustrate recurring configurations in how these support needs co-occurred. The analysis suggests a possibility of five configurations—Minimal Support, Utilitarian, Emo-Companion, Moderate-Low, and Resource-Oriented, as descriptive patterns. These patterns provide the insight that the affective and functional uses of AI have the tendency of co-occurrence in different ways. We then extend an exploratory analysis on how these user types might relate to participants' real-life support experiences and well-being. Our results indicate differences across user types. Users in the Minimal Support group report the highest levels of real-life support and well-being score, while users in the Emo-Companion group report the lowest. 

This study makes three contributions. First, it introduces a multidimensional AI-Support Scale, grounded in social support theory, that captures four dimensions of how users draw on AI: emotional/esteem, companionship, informational, and instrumental support. This addresses a gap in existing measures, which sometimes collapse AI use into a single emotional-versus-task-oriented distinction. Second, the result of this survey shows that users are not homogeneous---users seek support from generative AI differently in terms of their level of affective and functional needs. Third, it offers an exploratory insight for future research: users' affective and functional use of AI may be associated with their level of real-life support and well-being.  These contributions provide a structured way to understand differences across users, and highlight the importance of how generative AI system design could evaluate user affordance differently, particularly for those who heavily rely on AI to meet affective needs.
\section{Related Work}

\subsection{Human-AI Interaction through a Support-Based Needs Framework}
With the rise of Large Language Models (LLMs), conversational AI has become part of users' daily lives, taking on roles that involve social interaction and support, including social companionship \cite{chaturvedi2023companionship}, dating \cite{koike2023dating}, end-of-life planning \cite{utami2017endoflife}, and mental health treatment \cite{breso2016mentalhealth}. Much of this work is grounded in the CASA paradigm, which suggests that users instinctively apply interpersonal social behaviors to computer agents \cite{nass2000machines,brandtzaeg2021chatbots,meng2021disclose}. However, CASA primarily explains how users respond to AI as social actors, rather than what kinds of needs these interactions fulfill \cite{cornell2024limits}. To better characterize these needs, social support provides a structured framework for describing different types of help individuals receive through interaction \cite{cutrona1992spouse,schwarzer1991overview}.

Social support is commonly classified into multiple types, including informational, instrumental, emotional, esteem, and companionship-related support \cite{cutrona1992spouse,schwarzer1991overview,ta2020chatbot}. Informational and instrumental support are generally task-oriented, while emotional, esteem, and companionship support capture affective aspects of interaction \cite{cutrona1992spouse}. This distinction has been widely used to understand how different forms of support contribute to well-being and coping processes \cite{cohen1985buffering,chu2010socialsupport,schwarzer2007roles}. Prior work has applied this framework to AI, showing that conversational agents can provide emotional support comparable to human providers \cite{chin2023support,fulmer2018tess,kothgassner2019virtual,meng2021disclose,sinha2023wysa}, as well as informational support in task-oriented contexts \cite{brandtzaeg2021chatbots,skjuve2018chatbots,sohrabi2020covid}. However, existing studies often differentiate AI use into emotional and informational (or task-oriented) forms. While this distinction is useful, it does not capture how these forms of support may co-occur in practice, especially with the broad adoption of generative AI. As a result, it remains unclear how affective and functional needs are combined in actual AI use. In this work, we adopt this framework and focus on how these needs interact in users' engagement with AI.

\subsection{Current Human-AI Studies and the User Differences}
Existing work has moved beyond single exchanges to consider sustained relationships with AI. Research shows that AI can serve as a relationship partner, with users forming social bonds over time \cite{li2024love,pentina2023replika,purington2017alexa}. Several relationship types have been discussed, including companionship \cite{brandtzaeg2022myai,skjuve2021companion}, romantic relationships \cite{lei2024otome,leo2023gender}, and mentorship \cite{golan2024belong,sabnis2024teaching}.

Despite this, prior research often treats users as a relatively uniform audience, with limited attention to differences in user needs \cite{croes2021mitsuku,rapp2023collaboration,meng2021disclose}. For example, Croes and Antheunis \cite{croes2021mitsuku} focused on agents' relational capacity rather than user differences; Rapp et al.\ \cite{rapp2023collaboration} emphasized task-oriented interaction; and Meng and Dai \cite{meng2021disclose} examined AI communication patterns without accounting for user heterogeneity. Agent-centered explanations—such as anthropomorphic features \cite{cohn2024anthropomorphism,manzini2025code}, emotional expression \cite{liu2018sympathy}, and self-disclosure \cite{lee2020disclosure}—similarly focus on system properties rather than differences across users.

More importantly, existing work often distinguishes between emotional and task-oriented forms of AI use \cite{purington2017alexa,li2024love}. While this distinction is useful, it does not capture how users may simultaneously engage with multiple forms of support. Understanding these differences requires examining how affective and functional needs are combined in practice. The present study addresses this gap by mapping user patterns based on how these needs are combined, rather than treating them as separate or mutually exclusive categories. This gap motivates our second research question on how user patterns emerge from the combination of affective and functional needs.

\subsection{Real-Life Support and AI Support-Seeking}
Various individual and contextual factors may influence how users perceive AI-provided support. Perceived support is closely linked to the actual amount of support received \cite{gutierrez2024caregivers,robinson2021autism}, but also depends on how well that support matches an individual's needs \cite{melrose2015support}. Support-seeking intention further reflects an individual's willingness to initiate obtaining support \cite{stapley2023helpseeking}. 

Building on this perspective, real-life social support conditions can shape how users engage with AI-mediated support. AI companions have been shown to foster meaningful social bonds by mimicking real-world supportive interactions \cite{merrill2025chatbots,zhang2025companions}. Individuals experiencing loneliness are more likely to seek online relationships \cite{hood2018online}, and those with limited real-life support networks may form compensatory relationships with AI companions \cite{zhang2025companions}. These findings suggest that AI support-seeking is closely connected to users' broader social conditions. Well-being also plays an important role. It reflects both subjective happiness and perceived psychological state \cite{haver2015wellbeing}, and influences how individuals respond to stress and support \cite{tang2014jobstress}. Empirical work shows that lower well-being is associated with greater reliance on AI for emotional support \cite{zhang2025companions}. Overall, these findings indicate that users' engagement with AI is related to their real-life support level and well-being.
\section{Study Design and Analysis}

In this study, the term \emph{AI} refers to conversational AI. We employed a survey-based method to establish the factor structure of the AI-Support Scale and explore user profiles through cluster analysis.

To examine how users engage with different types of AI support, we reviewed existing social support scales, including the Social Support Behavior Code \cite{suhr1990support} and the MOS Social Support Survey \cite{sherbourne1991mos}. Because these scales were developed for human relationships, we adapted items from the Online Social Support Scale (OSSS; \cite{nick2018osss}), designed for online and computer-mediated environments. We organized measures around four support dimensions: affective (emotional/esteem), companionship, informational, and instrumental support. Cluster analysis was adopted to identify the potential user types. Additionally, we explored how affective and functional AI use relate to participants' real-life experiences: well-being, measured with the Short Warwick--Edinburgh Mental Well-Being Scale (SWEMWBS) \cite{haver2015wellbeing}, and real-life social support, measured with the Berlin Social Support Scales (BSSS) \cite{schwarzer1991overview}, specifically support needs, support-seeking behavior, and received support.

\subsection{Data Collection and Participants}
From July to August 2025, we collected survey data through an online platform (Wenjuan.com). After excluding responses with excessive missing data or failed manipulation checks, 1,090 valid responses were retained from 1,211 participants. Of these, 530 identified as male (48.6\%), 518 as female (47.5\%), 22 (2.0\%) as other, and 20 (1.8\%) preferred not to say. Participants were primarily young and middle-aged adults: 68.3\% were 18--34 years old, 23.4\% were 35--44 years old, and 8.3\% were 45 or older.

\subsection{Data Analysis}
To ensure measures captured distinct AI support dimensions, we conducted an exploratory factor analysis (EFA) on the initial item set. Factor loadings were examined following established guidelines \cite{hair2010multivariate}; items with loadings below .40 or substantial cross-loadings were removed.
Internal consistency was assessed using Cronbach's $\alpha$. To explore user engagement patterns, we conducted $k$-means cluster analysis on participants' scores across four support dimensions. After comparing multiple cluster solutions, we selected a five-cluster solution based on interpretability and stability. The $k$-means algorithm converged after six iterations. Four composite measures were calculated to analyze users' real-life conditions: Overall Well-Being (items 32--38), Need for Support (items 39--42), Support Seeking (items 43--46), and Received Support (items 47--50).
\section{Results}

\subsection{Factor Analysis}
The Kaiser-Meyer-Olkin (KMO) value was 0.96, exceeding the recommended
threshold of 0.90 and indicating excellent sampling adequacy. Bartlett's
test of sphericity was significant, $\chi^{2}(465) = 15{,}241.56$,
$p < .001$, confirming the data were appropriate for factor analysis.
Both statistics are reported in Table~\ref{tab:kmo}. The EFA supported a
four-factor solution (Table~\ref{tab:efa_loadings}), consistent with the
expected dimensions: emotional/esteem support, companionship, informational
support, and instrumental support. Five items (V14, V18, V19, V22, V27)
showed low loadings or substantial cross-loadings and were removed. The
final solution included 26 items with factor loadings ranging from 0.50
to 0.81, accounting for a substantial proportion of variance.

\begin{table}[t]
\caption{KMO and Bartlett's Test of Sphericity ($N = 1{,}090$)}
\label{tab:kmo}
\begin{center}
\begin{tabular}{lll}
  \toprule
  \textbf{Test} & \textbf{Value} & \textbf{$p$} \\
  \midrule
  Kaiser--Meyer--Olkin (KMO)    & .96                           & ---      \\
  Bartlett's Test of Sphericity & $\chi^{2}(465) = 15{,}241.56$ & $<.001$  \\
  \bottomrule
\end{tabular}
\end{center}
\end{table}

\begin{table}[t]
\caption{EFA Factor Loadings for the AI-Support Scale ($N = 1{,}090$).
Loadings ${<}$\,.40 omitted for clarity.}
\label{tab:efa_loadings}
\begin{center}
\begin{tabular}{lcccc}
  \toprule
  \textbf{Item} & \textbf{F1} & \textbf{F2} & \textbf{F3} & \textbf{F4} \\
                & \small{Emo.} & \small{Comp.} & \small{Info.} & \small{Inst.} \\
  \midrule
  V1  & .781 &      &      &      \\
  V2  & .668 &      &      &      \\
  V3  & .668 &      &      &      \\
  V4  & .781 &      &      &      \\
  V5  & .646 &      &      &      \\
  V6  & .759 &      &      &      \\
  V7  & .629 &      &      &      \\
  V8  & .751 &      &      &      \\
  V9  &      &      & .703 &      \\
  V10 &      &      & .662 &      \\
  V11 &      &      & .713 &      \\
  V12 &      &      & .751 &      \\
  V13 &      &      & .766 &      \\
  V15 &      &      & .711 &      \\
  V16 &      &      &      & .691 \\
  V17 &      &      &      & .702 \\
  V20 &      &      &      & .681 \\
  V21 &      &      &      & .503 \\
  V23 &      &      &      & .615 \\
  V24 &      &      &      & .782 \\
  V25 &      & .717 &      &      \\
  V26 &      & .778 &      &      \\
  V28 &      & .812 &      &      \\
  V29 &      & .607 &      &      \\
  V30 &      & .778 &      &      \\
  V31 &      & .745 &      &      \\
  \bottomrule
\end{tabular}
\end{center}
{\small \textit{Note.} F1 = Emotional/Esteem; F2 = Companionship;
F3 = Informational; F4 = Instrumental.}
\end{table}

\subsection{Reliability Analysis}

The overall item set showed excellent internal consistency ($\alpha = .93$).
Emotional/esteem support (eight items) reached $\alpha = .90$.
Social companionship showed $\alpha = .86$ after removing item V14.
Informational support achieved $\alpha = .83$ after excluding items V18,
V19, and V22. Instrumental support showed $\alpha = .86$ after removing
item V27. Most corrected item-total correlations exceeded .50, indicating
consistent alignment with respective dimensions. Full subscale statistics
are reported in Table~\ref{tab:reliability}.

\begin{table}[t]
\caption{Reliability of the AI-Support Subscales}
\label{tab:reliability}
\begin{center}
\begin{tabular}{lcc}
  \toprule
  \textbf{Subscale} & \textbf{Cronbach's $\alpha$} & \textbf{Item-Total $r$} \\
  \midrule
  Emotional \& Esteem & .901 & .62--.74 \\
  Companionship       & .858 & .62--.71 \\
  Informational       & .830 & .59--.69 \\
  Instrumental        & .863 & .68--.76 \\
  Overall Scale       & .928 & .52--.71 \\
  \bottomrule
\end{tabular}
\end{center}
\end{table}

\subsection{Cluster Analysis}

Inspection of the elbow plot of within-cluster sum of squares
(Fig.~\ref{fig:elbow} in the supplementary material) showed a marked decrease through
four to five clusters, after which additional clusters yielded diminishing
returns.
We selected a five-cluster solution. The $k$-means cluster analysis on
participants' scores across four support dimensions revealed distinct
profiles.

\textbf{Cluster~1: Moderate-Low Users} ($n = 235$) showed emotional
support around the average and slightly lower companionship, with clearly
below-average informational and instrumental support.
\textbf{Cluster~2: Utilitarian} ($n = 149$) had the lowest emotional and
companionship support, while relying somewhat on informational and
instrumental functions.
\textbf{Cluster~3: Resource-Oriented} ($n = 174$) was characterized by a
task-focused pattern, with the highest informational and instrumental
support, slightly elevated companionship, and relatively low emotional
support.
\textbf{Cluster~4: Emo-Companion} ($n = 378$) emphasized relational
aspects, scoring highest in both emotional and companionship support,
while informational and instrumental support were moderately above the
mean.
\textbf{Cluster~5: Minimal Support} ($n = 154$) represented a low-support
profile, with all four dimensions well below the sample average.

\begin{figure*}[t]
  \centering
  \includegraphics[width=0.98\textwidth]{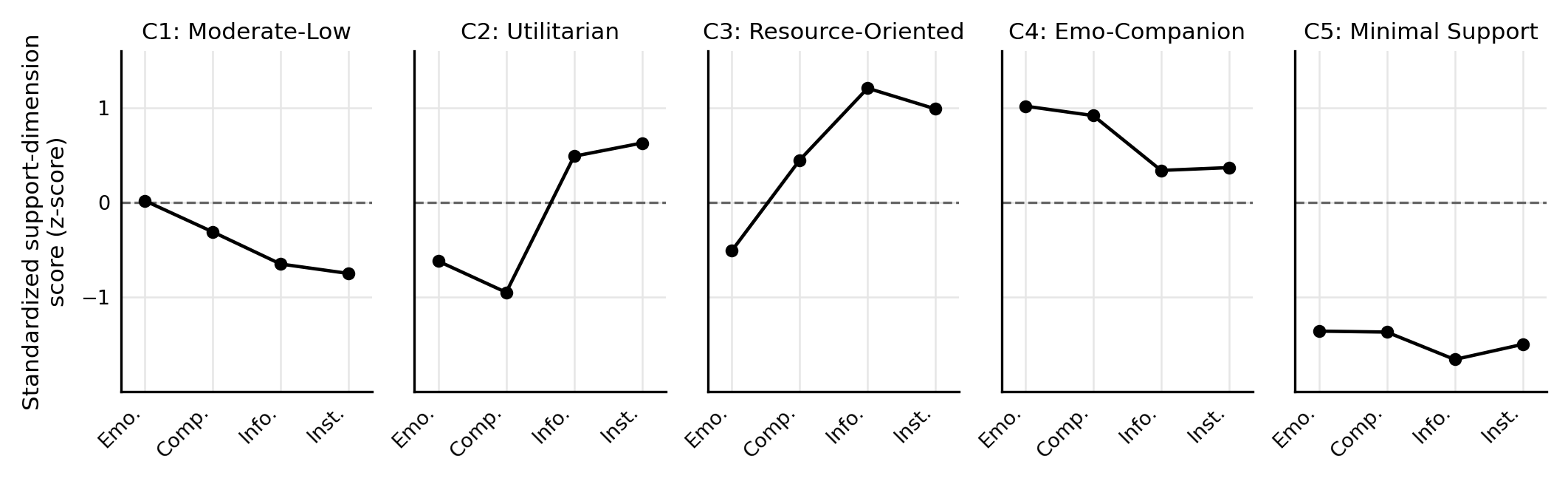}
  \caption{Standardized mean scores of the five user clusters on the AI-Support subscales. Each panel represents one cluster; the horizontal axis shows emotional/esteem (Emo.), social companionship (Comp.), informational (Info.), and instrumental (Inst.) support. Dashed lines indicate the sample mean ($z = 0$).}
  \label{fig:cluster_means}
\end{figure*}

These findings suggest five user profiles, highlighting heterogeneity of engagement patterns. Cluster means are presented in Table~\ref{tab:cluster_means} and visualized in Fig.~\ref{fig:cluster_means}.
\begin{table}[t]
\caption{Cluster Means for Perceived AI Support by Support Type}
\label{tab:cluster_means}
\begin{center}
\begin{tabular}{llrrrr}
  \toprule
  Cluster & $n$ & Emo. & Comp. & Info. & Inst. \\
  \midrule
  1 (Moderate-Low)     & 235 &  \phantom{$-$}0.02 & $-$0.31 & $-$0.65 & $-$0.75 \\
  2 (Utilitarian)      & 149 & $-$0.62 & $-$0.95 &  \phantom{$-$}0.49 &  \phantom{$-$}0.63 \\
  3 (Resource-Oriented)& 174 & $-$0.51 &  \phantom{$-$}0.45 &  \phantom{$-$}1.21 &  \phantom{$-$}0.99 \\
  4 (Emo-Companion)    & 378 &  \phantom{$-$}1.02 &  \phantom{$-$}0.92 &  \phantom{$-$}0.34 &  \phantom{$-$}0.37 \\
  5 (Minimal Support)  & 154 & $-$1.36 & $-$1.37 & $-$1.66 & $-$1.50 \\
  \bottomrule
\end{tabular}
\end{center}
{\small \textit{Note.} Values are standardized ($z$-scores). Emo.\
= Emotional/Esteem; Comp.\ = Social Companionship; Info.\
= Informational; Inst.\ = Instrumental.}
\end{table}

\subsection{Real-Life Support Experience and Well-Being}
The Resource-Oriented group reported moderately high scores in well-being
($M = 3.18$) and support outcomes, above the Utilitarian
and Moderate-Low groups but below the Minimal Support group. The
Utilitarian group demonstrated mid-range outcomes (well-being and support
means between 2.41--2.78). The Moderate-Low group reported relatively low
scores, though still higher than the Emo-Companion group.
Cluster means and standard deviations for all outcome variables are
reported in Table~\ref{tab:cluster_outcomes}.
\begin{table}[t]
\caption{Means and Standard Deviations of Outcome Variables by Cluster}
\label{tab:cluster_outcomes}
\begin{center}
\setlength{\tabcolsep}{3.5pt}
\footnotesize
\begin{tabular}{lcccc}
  \toprule
  \textbf{Cluster} & \textbf{Well-being} & \textbf{Need} & \textbf{Seeking} & \textbf{Received} \\
                   & $M$ (SD)            & $M$ (SD)      & $M$ (SD)         & $M$ (SD)          \\
  \midrule
  Minimal Support   & 3.60 (0.61) & 3.08 (0.73) & 3.68 (0.65) & 3.58 (0.79) \\
  Moderate-Low      & 2.41 (0.79) & 2.19 (0.83) & 2.47 (0.81) & 2.37 (0.81) \\
  Utilitarian       & 2.73 (0.73) & 2.41 (0.82) & 2.78 (0.90) & 2.73 (0.84) \\
  Resource-Oriented & 3.18 (0.66) & 2.96 (0.68) & 3.36 (0.69) & 3.22 (0.69) \\
  Emo-Companion     & 1.73 (0.61) & 1.55 (0.47) & 1.82 (0.69) & 1.76 (0.66) \\
  \bottomrule
\end{tabular}
\end{center}
\end{table}
From the result, the Minimal Support group demonstrates the most favorable
real-life outcomes, whereas the Emo-Companion group occupies the least
advantageous position. The result is only an indicator of possible relations between subjective well-being and human-AI interaction. No correlation analysis was conducted due to the instability of the number of clusters.
\section{Discussion \& Conclusion}
Acknowledging the conversational and relational nature of human–AI interaction (HAII), this study examines HAII through the user-needs-fulfillment scope using social support framework \cite{cutrona1992spouse,schwarzer2007roles,pan2025framework}. Critically, Cutrona and Suhr treat people's affective and functional needs as analytically separable but empirically co-occurring dimensions — not mutually exclusive categories. Instead of separating users into binary groups, our findings show that users simultaneously engage with both affective (e.g., emotional reassurance and companionship) and functional (e.g., informational and instrumental) uses. A single AI interaction can simultaneously engage multiple support dimensions, and we developed a scale to capture this co-occurrence nature. This extends prior work by understanding users from a holistic user needs centered angle, compared to the commonly used measuring and differentiating HAII with binary emotional-versus-functional classifications.  While some groups align with prior distinctions between emotional and task-oriented AI use \cite{purington2017alexa,li2024love}, our results suggest results beyond the binary view. 

Further exploration of real-life outcome variables indicates a potential that people's real-life support and well-being varies by their AI engagement type. As the group with least demand of using AI, The Minimal Support users are showing the highest level of well-being condition and received real-life social support. Though this study didn't give the answer of the significant correlations, either the causal effects among these variables, the insight is corresponding to the ``compensatory hypothesis" in the social support literature. ``Compensatory hypothesis" holds that when real-life interpersonal support is sufficient, individuals are less likely to turn to alternative sources \cite{cutrona1990matching}. Meanwhile, for individuals with high levels of well-being, their use of AI may serve more as a ``supplementary resource'' that provides added value \cite{garcia2024role}.

This study has several limitations. First, the data were collected from participants in China, and patterns of AI use and affective reliance may differ across cultural contexts where norms around emotional expression, social support, and technology use vary. Therefore, the generalizability of these findings to other populations should be interpreted with caution. 

Second, the use of cluster analysis introduces methodological constraints. The number and structure of clusters are not fixed and may vary depending on the dataset, feature selection, and clustering method. Although techniques such as the elbow method were used to guide the selection, the resulting clusters still involve subjective decisions. This limit also highlights the study contribution in how users' perceived AI effects co-occur in different proportions, forming structured yet continuous patterns of AI engagement without imposing any categorical boundaries. 

Future research should investigate whether the balance between affective and functional use reflects stable user traits or dynamic adaptation to changing life conditions. The current findings suggest that users differ in how they combine these needs, but it is unclear whether these patterns remain consistent over time or shift in response to personal circumstances, such as changes in social support, stress, or life transitions.

Additionally, the observed differences in patterns of AI use raise important design questions regarding user-specific adoption and risks. Users who rely more heavily on affective use may have different expectations regarding emotional responsiveness, availability, and continuity of interaction \cite{chin2023support,fulmer2018tess}, while users with more functional use may prioritize efficiency and task support \cite{brandtzaeg2021chatbots,skjuve2018chatbots}. Future work should examine how users' different expectations on AI supports are related to their sensitivity and affordance to AI adoption, such as AI over-reliance or anxiety from AI withdrawal, and how AI systems can be designed to mitigate potential psychological risks for vulnerable users. 
\section*{Ethical Impact Statement}

\textbf{Human Subjects Research.}
All study procedures were reviewed and granted exempt status by the
Massachusetts Institute of Technology Institutional Review Board
(Exempt ID: E-7025). Participation was fully voluntary. Prior to
completing the survey, all participants received a digital informed
consent form clearly explaining the study objectives, procedures, data
storage and privacy practices, their right to withdraw at any time
without penalty, and the absence of foreseeable risks. Completion of the
survey constituted consent. Participants were compensated at a rate consistent with local market norms for online survey tasks via the Wenjuan.com platform. No
deception was used in this study. The survey was administered in Chinese
and adapted for the cultural context of mainland China.

\textbf{Potential Harms.}
This study does not involve deception. The survey asked participants
about their AI support-seeking behaviors; no
intervention was administered and no sensitive clinical data were
collected.  Our findings should inform protective and equitable AI design rather than targeted
commercial exploitation. The study sample was drawn entirely from
mainland China, and findings should not be generalized to other
demographic, cultural, or geographic populations without appropriate
validation. No surveillance-enabling or rights-limiting applications of
this research are intended or endorsed.

\textbf{Bias and Demographic Considerations.}
All participants were adults recruited through an online platform
(Wenjuan.com), which may systematically over-represent younger, more
digitally literate, and urban individuals. The sample's demographic
distribution---predominantly 18--44 years old, with gender not evenly
distributed---may limit generalizability to older populations, rural
users, or individuals with limited digital access. Additionally, AI
usage norms, social support structures, and psychological responses to
AI interaction may differ substantially across cultural contexts. The
five-cluster typology derived here should be interpreted as contextually
situated rather than universal. Future work should replicate these
findings across culturally and demographically diverse samples.

\textbf{Generalizability Limits.}
This study was conducted over a specific two-month window (July--August
2025) and reflects users' perceptions of AI at a particular moment in
the technology's development. As conversational AI capabilities evolve
rapidly, the specific support profiles and cluster structures identified
here may shift.

\textbf{Data and Artifact Availability.}
The survey instrument and anonymized aggregate summary statistics will be
made available upon reasonable request from the authors. Individual-level
data cannot be released publicly due to IRB restrictions and participant
privacy commitments. No new AI systems, models, or datasets requiring
independent ethical review were produced as part of this work.

\bibliographystyle{IEEEtran}
\bibliography{citation}

\clearpage
\onecolumn
\section*{Supplementary Material}
\subsection*{Cluster-Count Selection}
This supplement reports the within-cluster sum of squares (WSS) diagnostic
used to select the number of $k$-means clusters in the main paper.
As shown in Fig.~\ref{fig:elbow}, the WSS decreases sharply through four to
five clusters and flattens thereafter, supporting the five-cluster solution
reported in the main text.
\setcounter{figure}{0}
\renewcommand{\thefigure}{S\arabic{figure}}
\begin{figure}[h]
  \centering
  \includegraphics[width=0.65\textwidth]{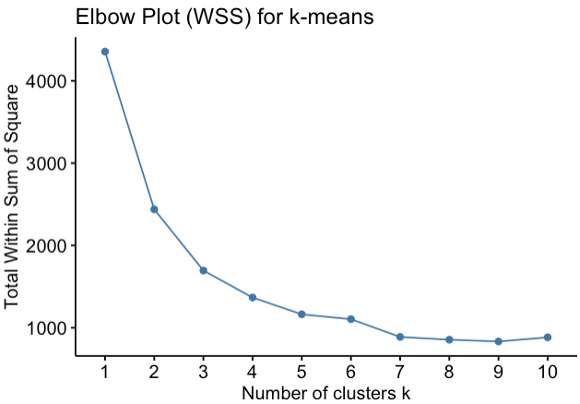}
  \caption{Elbow plot of within-cluster sum of squares (WSS) for $k = 1$--$10$ clusters. The slope decreases sharply through four or five clusters before flattening, supporting $k = 5$ as the selected solution.}
  \label{fig:elbow}
\end{figure}
\end{document}